\documentclass[aoas,MSNbibl,nameyear,seceqn,dvips]{arximspdf}
\usepackage{mathbh}
\usepackage{graphicx}

\doi{10.1214/14-AOAS767} 
\volume{8}
\issue{4}
\pubyear{2014}
\firstpage{2247}
\lastpage{2267}
\docsubty{FLA}

\makeatletter
\newcommand{\rrvert}{\vert}
\newcommand{\llvert}{\vert}
\makeatother

\begin{document}
\begin{frontmatter}

\title{Voronoi residual analysis of spatial point process models with
applications to California earthquake~forecasts}
\runtitle{Voronoi residuals for point processes}

\begin{aug}
\author[A]{\fnms{Andrew} \snm{Bray}\thanksref{m1}\ead[label=e1]{andrew.bray@gmail.com}},
\author[B]{\fnms{Ka} \snm{Wong}\thanksref{m2}\ead[label=e2]{kawong@google.com}},
\author[C]{\fnms{Christopher D.} \snm{Barr}\thanksref{m3}\ead[label=e3]{cdbarr@gmail.com}}\\
\and
\author[D]{\fnms{Frederic Paik} \snm{Schoenberg}\corref{}\thanksref{m4}\ead[label=e4]{frederic@stat.ucla.edu}}
\runauthor{Bray, Wong, Barr and Schoenberg}
\affiliation{University of Massachusetts, Amherst\thanksmark{m1},
Google\thanksmark{m2},
Yale University\thanksmark{m3}\\
and
University of California, Los Angeles\thanksmark{m4}}
\address[A]{A. Bray\\
Department of Mathematics and Statistics\\
University of Massachusetts, Amherst\\
Lederle GR Tower Box 34515\\
Amherst, Massachusetts 01003-9305\\
USA\\
\printead{e1}}
\address[B]{K. Wong\\
Google\\
3465 Deering St.\\
Oakland, California 94601\hspace*{46.5pt}\\
USA\\
\printead{e2}}
\address[C]{C. D. Barr\\
Yale School of Management\\
165 Whitney Ave.\\
New Haven, Connecticut 06511\\
USA\\
\printead{e3}}
\address[D]{F. P. Schoenberg\\
Department of Statistics\\
University of California, Los Angeles\\
8125 Math-Science Building\\
Los Angeles, California 90095-1554\\
USA\\
\printead{e4}} 
\end{aug}

\received{\smonth{7} \syear{2013}}
\revised{\smonth{7} \syear{2014}}

%
\begin{abstract}
Many point process models have been proposed for describing and
forecasting earthquake occurrences in
seismically active zones such as California, but the problem of how
best to compare and evaluate the
goodness of fit of such models remains open. Existing techniques
typically suffer from low power,
especially when used for models with very volatile conditional
intensities such as those used to
describe earthquake clusters. This paper proposes a new residual
analysis method for spatial or
spatial--temporal point processes involving inspecting the differences
between the modeled
conditional intensity and the observed number of points over the
Voronoi cells generated by
the observations. The resulting residuals can be used to construct
diagnostic methods of
greater statistical power than residuals based on rectangular grids.

Following an evaluation of performance using simulated data, the
suggested method is used to
compare the Epidemic-Type Aftershock Sequence (ETAS) model to the
Hector Mine earthquake catalog. The proposed residuals indicate that
the ETAS model with uniform background rate appears to slightly but
systematically underpredict seismicity along the fault and to
overpredict seismicity in along the periphery of the fault.
\end{abstract}

%
\end{frontmatter}

\section{Introduction}\label{sec1}

Considerable effort has been expended recently to assess and compare
different space--time models for forecasting earthquakes in seismically
active areas such as Southern California.
Notable among these efforts were the development of the Regional
Earthquake Likelihood Models (RELM) project [\citet{Field2007}] and its
successor, the Collaboratory for the Study of Earthquake Predictability
(CSEP) [\citet{Jordan2006}].
The RELM project was initiated to create a variety of earthquake
forecast models for seismic hazard assessment in California.
Unlike previous projects that addressed the assessment of models for
seismic hazard, the RELM participants decided to adopt many competing
forecasting models and to rigorously and prospectively test their
performance in a dedicated testing center [\citet{Schorlemmer2007}].
With the end of the RELM project, the forecast models became available
and the development of the testing center was done within the scope of CSEP.
Many point process models, including multiple variants of the
Epidemic-Type Aftershock Sequence (ETAS) models of \citet
{Ogata1998} have now been proposed and are part of RELM and CSEP,
though the problem of how to compare and evaluate the goodness of fit
of such models remains quite open.

In RELM, a community consensus was reached that all entered models be
tested with certain tests, including the Number or N-test that compares
the total forecasted rate with the observation, the Likelihood or
L-test that assesses the quality of a forecast in terms of overall
likelihood, and the Likelihood-Ratio or R-test that assesses the
relative performance of two forecast models compared with what is
expected under one proposed model [\citet
{Jackson1999,Schorlemmer2007b,Zechar2010,Rhoades2011,Zechar2013}].
However, over time several drawbacks of these tests were discovered
[\citet{Schorlemmer2010}] and the need for more powerful tests became clear.
The N-test and L-test simply compare the quantiles of the total numbers
of events in each bin or likelihood within each bin to those expected
under the given model, and the resulting low-power tests are typically
unable to discern significant lack of fit unless the overall rate of
the model fits extremely poorly. Further, even when the tests do reject
a model, they do not typically indicate where or when the model fits
poorly, or how it could be improved. Meanwhile, the number of proposed
spatial--temporal models for earthquake occurrences has grown, and the
need for discriminating which models fit better than others has become
increasingly important. Techniques for assessing goodness of fit are
needed to pinpoint where existing models may be improved, and residual
plots, rather than numerical significance tests, seem preferable for
these purposes.

This paper proposes a new form of residual analysis for assessing the
goodness of fit of spatial point process models. The proposed method
compares the normalized observed and expected numbers of points over
Voronoi cells generated by the observed point pattern.
The method is applied here in particular to the examination of a
version of the ETAS model originally proposed by \citet
{Ogata1998}, and its goodness of fit to a sequence of 520 $M \geq3$
Hector Mine earthquakes occurring between October 1999 and December 2000.
In particular, the Voronoi residuals indicate that assumption of a
constant background rate $\rho$ in the ETAS model results in excessive
smoothing of the seismicity and significant underprediction of
seismicity close to the fault line.


Residual analysis for a spatial point process is typically performed by
partitioning the space on which the process is observed into a regular
grid and computing a residual for each pixel.
That is, one typically examines aggregated values of a residual process
over regular, rectangular grid cells.
Alternatively, residuals may be defined for every observed point in the
pattern, using a metric such as deviance, as suggested in \citet
{Lawson1993}.
Various types of residual processes were proposed in \citet
{Baddeley2005b} and discussed in \citet{Baddeley2008} and
\citet{Clements2011}.
The general form of these aggregated residual measures is a
standardized difference between the number of points occurring and the
number expected according to the fitted model, where the
standardization may be performed in various ways.
For instance, for Pearson residuals,
one weights the residual by the reciprocal of the square root of the
intensity, in analogy with Pearson residuals in the context of linear models.
\citet{Baddeley2005b} propose smoothing the residual field using a
kernel function instead of simply aggregating over pixels; in practice,
this residual field is typically displayed over a rectangular grid and
is essentially equivalent to a kernel smoothing of aggregated pixel residuals.
\citet{Baddeley2005b} also propose scaling the residuals based on
the contribution of each pixel to the total pseudo-loglikelihood of the
model, in analogy with score statistics in generalized linear modeling.
Standardization is important for both residual plots and
goodness-of-fit tests, since otherwise plots of the residuals will tend
to overemphasize deviations in pixels where the rate is high.
Behind the term \textit{Pearson residuals} lies the implication [see,
e.g., the error bounds in Figure~7 of \citet{Baddeley2005b}] that
these standardized residuals should be approximately standard normally
distributed, so that the squared residuals, or their sum, are
distributed approximately according to Pearson's $\chi^2$-distribution.

The excellent treatment of Pearson residuals and other scaled residuals
by \citet{Baddeley2005b}, the thorough discussion of their
properties in \citet{Baddeley2008}, their use for formal inference
using score and pseudo-score statistics as described in \citet
{Baddeley2011}, and the fact that such residuals extend so readily to
the case of spatial--temporal point processes may suggest that the
problem of residual analysis for such point processes is generally
solved. In practice, however, such residuals, when examined over a
fixed rectangular grid, tend to have two characteristics that can limit
their effectiveness:
\begin{longlist}[II.]
\item[I.] When the integrated conditional intensity (i.e., the number of
expected points) in a pixel is very small, the distribution of the
residual for the pixel becomes heavily skewed.

\item[II.] Positive and negative values of the residual process within a
particular cell can cancel each other out.
\end{longlist}

Since Pearson residuals are standardized to have mean zero and unit (or
approximately unit) variance under the null hypothesis that the modeled
conditional intensity is correct [see \citet{Baddeley2008}], one
may inquire whether the skew of these residuals is indeed problematic.
Consider, for instance, the case of a planar Poisson process where the
estimate of the intensity $\lambda$ is exactly correct, that is, $\hat
\lambda(x,y) = \lambda(x,y)$ at all locations, and where one elects to
use Pearson residuals on pixels. Suppose that there are several pixels
where the integral of $\lambda$ over the pixel is roughly $0.01$. Given
many of these pixels, it is not unlikely that at least one of them will
contain a point of the process. In such pixels, the raw residual will
be $0.99$, and the standard deviation of the number of points in the
pixel is $\sqrt{0.01} = 0.1$, so the Pearson residual is $9.90$. This
may yield the following effects: (1) such Pearson residuals may
overwhelm the others in a visual inspection, rendering a plot of the
Pearson residuals largely useless in terms of evaluating the quality of
the fit of the model, and (2) conventional tests based on the normal
approximation may have grossly incorrect $p$-values, and will commonly
reject the null model even when it is correct. Even if one adjusts for
the nonnormality of the residual and instead uses exact $p$-values
based on the Poisson distribution, such a test applied to any such
pixel containing a point will still reject the model at the
significance level of $0.01$.

These situations arise in many applications, unfortunately. For
example, in modeling earthquake occurrences, typically the modeled
conditional intensity is close to zero far away from known faults or
previous seismicity, and in the case of modeling wildfires, one may
have a modeled conditional intensity close to zero in areas far from
human use or frequent lightning, or with vegetation types that do not
readily support much wildfire activity [e.g., \citet
{Johnson2001,Malamud2005,Keeley2009,Xu2011}].

These challenges are a result of characteristic I above, and one
straightforward solution would be to enlarge the pixel size such that
the expected count in each cell is higher. While this would be
effective in a homogeneous setting, in the case of an inhomogeneous
process it is likely that this would induce a different problem: cells
that are so large that even gross misspecification within a cell may be
overlooked, and thus the residuals will have low power. This is the
problem of characteristic~II. When a regular rectangular grid is used
to compute residuals for a highly inhomogenous process, it is generally
impossible to avoid either highly skewed residual distributions or
residuals with very low power. These problems have been noted by
previous authors, though the important question of how to determine
appropriate pixel sizes remains open [\citet{Lawson2005}].

Note that, in addition to Pearson residuals and their variants, there
are many other goodness-of-fit assessment techniques for spatial and
spatial--temporal point processes [\citet{Bray2013}].
Examples include rescaled residuals [\citet{Meyer1971,Schoenberg1999}]
and superthinned residuals [\citet{Clements2012}], which involve
transforming the observed points to form a new point process that is
homogeneous Poisson under the null hypothesis that the proposed model
used in the transformation is correct.
There are also functional summaries, such as the weighted version [\citet
{Baddeley2000,Veen2005}] of Ripley's $K$-function [\citet{Ripley1976}],
where each point is weighted according to the inverse of its modeled
conditional intensity so that the resulting summary has conveniently
simple properties under the null hypothesis that the modeled
conditional intensity is correct, as well as other similarly weighted
numerical and functional summaries such as the weighted R/S statistic
and weighted correlation integral [\citet{Adelfio2009}].
As noted in \citet{Bray2013}, all of these methods can have
serious deficiencies compared to the easily interpretable residual
diagrams, especially when it comes to indicating spatial or
spatial--temporal locations where a model may be improved.

This paper proposes a new form of residual diagram based on the Voronoi
cells generated by tessellating the observed point pattern. The
resulting partition obviates I and II above by being adaptive to the
inhomogeneity of the process and generating residuals that have an
average expected count of 1 under the null hypothesis.

For an \hyperref[sec1]{Introduction} to point processes and their intensity functions,
the reader is directed to \citet{Daley1988}. Throughout this paper
we are assuming that the point processes are simple and that the
observation region is a
complete separable metric space equipped with Lebesgue measure, $\mu
$. Note that we are not emphasizing the distinction between conditional
and Papangelou intensities, as the methods and results here are
essentially equivalent for spatial and spatial--temporal point processes.

This paper is organized as follows. Section~\ref{sec2} describes Voronoi
residuals and discusses their properties. Section~\ref{sec3} demonstrates the
utility of Voronoi residual plots. The simulations shown in Section~\ref{sec4}
demonstrate the advantages of Voronoi residuals over conventional
pixel-based residuals in terms of statistical power. In Section~\ref{sec5} we
apply the proposed Voronoi residuals to examine the fit of the ETAS
model with uniform background rate to a sequence of Hector Mine
earthquakes from October 1999 to December 2000, and show that despite
generally good agreement between the model and data, the ETAS model
with uniform background rate appears to slightly but systematically
underpredict seismicity along the fault line and to overpredict
seismicity in certain locations along the periphery of the fault line,
especially at the location 35 miles east of Barstow, CA.

%
%
\section{Voronoi residuals}\label{sec2}\label{secvoronoiresiduals}

A Voronoi tessellation is a partition of the metric space on which a
point process is defined into convex polygons, or \textit{Voronoi cells}.
Specifically, given a spatial or spatial--temporal point pattern $N$,
one may define its corresponding \textit{Voronoi tessellation} as follows:
for each point $\tau_i$ of the point process, its corresponding cell
$C_i$ is the region consisting of all locations which are closer to
$\tau_i$ than to any other point of $N$. The Voronoi tessellation is
the collection of such cells. See, for example, \citet{Okabe2000}
for a thorough treatment of Voronoi tessellations and their properties.

Given a model for the conditional intensity of a spatial or space--time
point process, one may construct residuals simply by evaluating the
residual process over cells rather than over rectangular pixels, where
the cells comprise the Voronoi tessellation of the observed spatial or
spatial--temporal point pattern. We will refer to such residuals as \textit{Voronoi residuals}.

An immediate advantage of Voronoi residuals compared to conventional
pixel-based methods is that the partition is entirely automatic and
spatially adaptive. This leads to residuals with a distribution that
tends to be far less skewed than pixel-based methods. Indeed, since
each Voronoi cell has exactly one point inside it by construction, the
raw Voronoi residual for cell $i$ is given by
%
\begin{eqnarray}\label{eq1}
\hat r_i &:=& 1 - \int
_{C_i} \hat\lambda\,\mathrm{d}\mu
\nonumber\\[-8pt]\\[-8pt]\nonumber
&=& 1 - \llvert C_i\rrvert\bar\lambda,
\end{eqnarray}
where $\bar\lambda$ denotes the mean of the proposed model, $\hat
\lambda$, over $C_i$. This raw residual can be scaled in various ways,
as is well addressed 
by \citet{Baddeley2005b}.
Note that when $N$ is a homogeneous Poisson process, the sizes
$\llvert C_i\rrvert $ of the cells are approximately gamma
distributed. Indeed, for a homogeneous Poisson process, the expected
area of a Voronoi cell is equal to the reciprocal of the intensity of
the process [\citet{Meijering1953}], and simulation studies have shown
that the area of a typical Voronoi cell is approximately gamma
distributed [\citet{Hinde1980,Tanemura2003}]; these properties continue
to hold approximately in the inhomogeneous case provided that the
conditional intensity is approximately constant near the location in
question [\citet{Barr2010,Barr2012}].

%
\begin{figure}[b]

\includegraphics{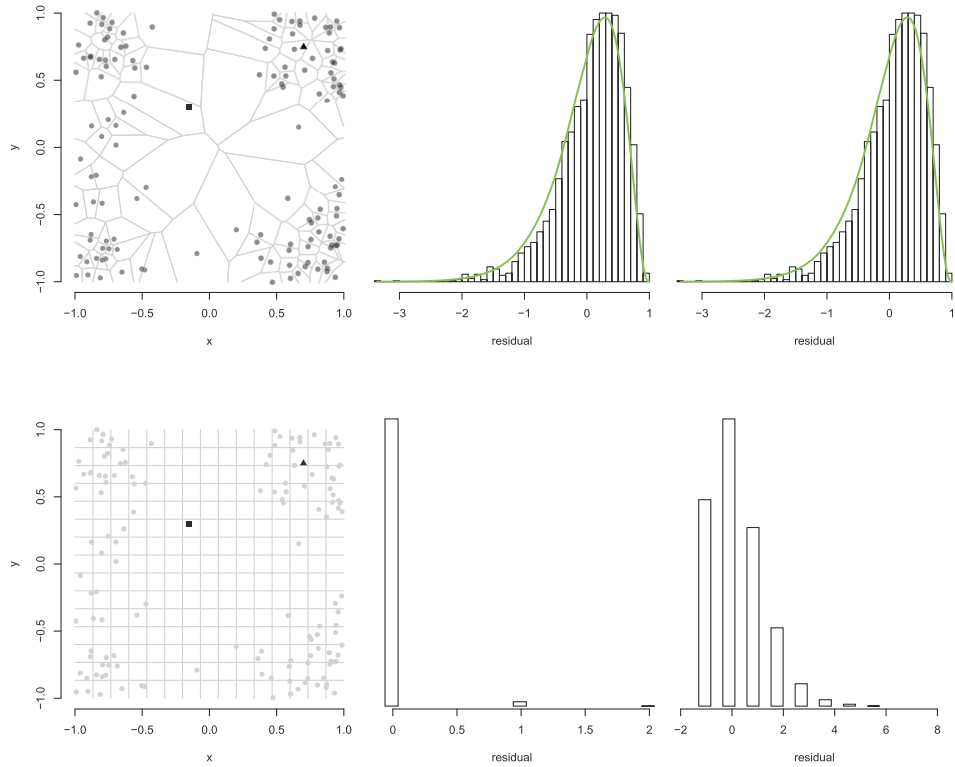}

\caption{Residual distributions under the null hypothesis based on a
Voronoi tessellation (top panels) and a pixellated grid (bottom
panels). The underlying point process is Poisson with intensity $\lambda
(x,y) = 200 x^2\llvert y\rrvert $. The middle panels show results at
location $\blacksquare$ where $\lambda= 1.35$; the right panels show
results at location $\blacktriangle$ where $\lambda= 73.5$. The
distribution (\protect\ref{eq2}) is overlaid for the top middle and top right plots.}\label{vorPixDists}
\end{figure}

The raw Voronoi residual in (\ref{eq1}) will therefore tend to be
distributed approximately like a modified gamma random variable. More
specifically, the second term, $\llvert C_i\rrvert \bar\lambda$,
referred to in the stochastic geometry literature as the reduced area,
is well approximated by a two-parameter gamma distribution with a rate
of 3.569 and a shape of 3.569 [\citet{Tanemura2003}]. The distribution
of the raw residuals is therefore approximated by
%
\begin{equation}
r \sim1 - X; \qquad X \sim\Gamma(3.569, 3.569). \label{eq2}
\end{equation}
By contrast, for pixels over which the integrated conditional intensity
is close to zero, the conventional raw residuals are approximately
Bernoulli distributed.

The exact distributions of the Voronoi residuals are generally quite
intractable due to the fact that the cells themselves are random, but
approximations can be made using simulation. Consider the point process
defined by the intensity function $\lambda(x,y) = 200x^2\llvert
y\rrvert $ on the subset $S = [-1,1] \times[-1,1]$. Figure~\ref{vorPixDists} presents a realization of the process along with the
corresponding Voronoi tessellation (top panels) and a regular
rectangular pixel grid (bottom panels). Two locations in $S$ were
selected for investigation: one with relatively high intensity, the
other with relatively low intensity. Residual distributions were
simulated by generating 5000 point patterns from the model, identifying
the pixel/cell occupied by the location of interest, then calculating
the difference between the number of observed points and the number
expected under the same model.

It can be seen from Figure~\ref{vorPixDists} that the distribution of
Voronoi residuals under the null hypothesis is well approximated by
distribution (\ref{eq2}) at both the high intensity and low intensity
locations. By comparison, the distribution of pixel residuals is that
of a Poisson distributed variable with intensity $\int_{G_i}
\hat{\lambda} \,\mathrm{d}\mu$, for pixel $G_i$, centered to have mean
zero. At the location where the intensity is high, this distribution is
moderately skewed, but for the low intensity location the distribution
becomes extremely skewed to the point of being effectively a two-valued
random variable.

\begin{figure}

\includegraphics{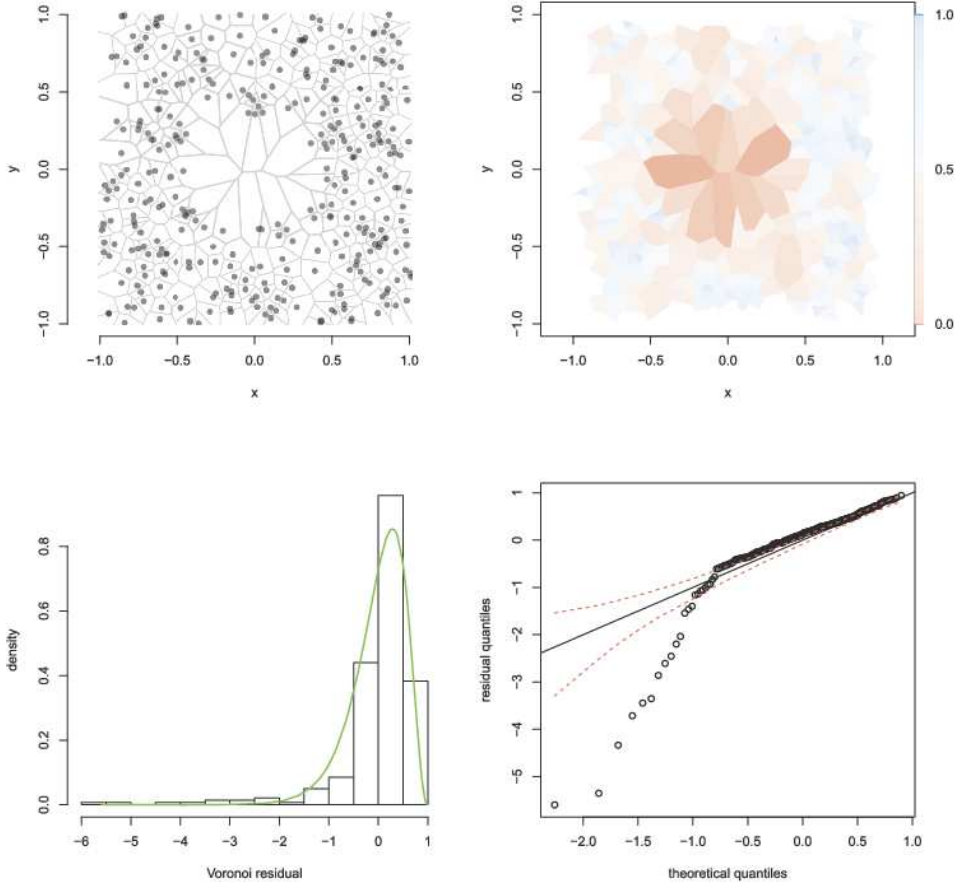}

\caption{Simulated Poisson process with intensity $\lambda(x,y) = 100
{\mathbf1}_{\{\llvert x\rrvert,\llvert y\rrvert > 0.35\}}$ with
Voronoi tessellation overlaid (top left), Voronoi residual plot for
this simulation using a proposed intensity of $\lambda(x,y) = 100$ (top
right), histogram of the Voronoi residuals, with the density of the
reference distribution (\protect\ref{eq2}) overlaid (bottom left),
quantile plot of the Voronoi residuals with respect to distribution~(\protect\ref{eq2}), with pointwise 95\% confidence limits obtained via
simulation of the proposed model (bottom right). The color scale of the
Voronoi residual plot is $\Phi^{-1}\{F(r)\}$, where $F$ is the
distribution function of (\protect\ref{eq2}). Tiles intersecting the
boundary of the space are ignored.}
\label{overestimating}
\end{figure}

In addition to distribution (\ref{eq2}) being a good approximation for
the distribution of the Voronoi residuals at a given location across
many realizations, it is also a close approximation for the
distribution of all residuals for a single realization (see, e.g., the
lower left plot in Figure~\ref{overestimating}). It is important to
note that such residuals are not strictly independent of one another
due to the nature of the tessellation, however, our assessment is that
this dependence is fairly minor. See the discussion for additional
comments on independence.

%

%
%
\section{Voronoi residual plots}\label{sec3}

In this section the utility of Voronoi residuals will be demonstrated
using a series of simulations of spatial Poisson processes. The
simulations are random samples from a specified \emph{generating
model}. These simulations are then modeled, correctly or incorrectly,
by a \emph{proposed model}. Voronoi residuals are then computed and
used to assess the degree to which the proposed model agrees with simulations.

\subsection{Correctly specified model}\label{sec3.1}
We first consider the simplest case, where the proposed model is the
same as the generating model. As a result, one expects residuals that
are spatially unstructured and relatively small in magnitude, that is,
the only variation should be due to sampling variability.

%
\begin{figure}

\includegraphics{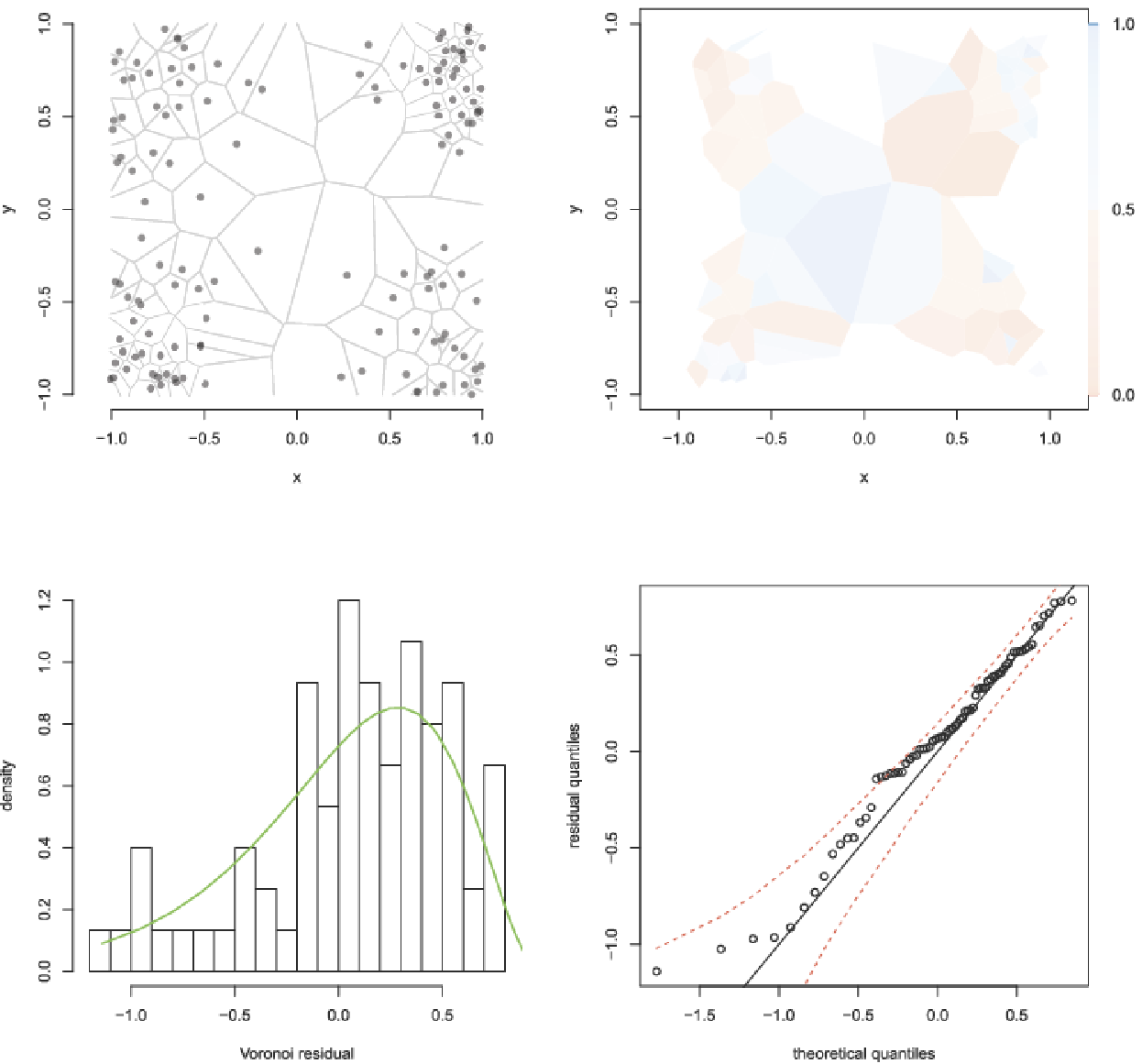}

\caption{Simulated Poisson process with intensity $\lambda(x,y) =
200x^2\llvert y\rrvert $ with Voronoi tessellation overlaid (top
left), Voronoi residual plot of this simulation (top right), histogram
of the Voronoi residuals, with the density of the reference
distribution (\protect\ref{eq2}) overlaid (bottom left), quantile plot
of the Voronoi residuals with respect to the distribution (\protect\ref
{eq2}), with pointwise 95\% confidence limits obtained via simulation
(bottom right). The color scale of the Voronoi residual plot in the top
right is $\Phi^{-1}\{F(r)\}$, where $F$ is the distribution function of
(\protect\ref{eq2}). Tiles intersecting the boundary of the space are
ignored, as the distribution of these tile areas may differ
substantially from the gamma distribution.}
\label{correctModel}
\end{figure}

Figure~\ref{correctModel} shows a simulation of a spatial Poisson
process with intensity $\lambda(x,y) = 200x^2\llvert y\rrvert $ on
the subset $S = [-1,1] \times[-1,1]$, along with its corresponding
Voronoi tessellation.

In the Voronoi residual plot in the top right panel of Figure~\ref
{correctModel}, each tile is shaded according to the value of the
residual under the distribution function of the modified gamma
distribution (\ref{eq2}). The resulting $p$-values are then mapped to the
color scale using an inverse Normal transformation. Thus, brightly
shaded red areas indicate unusually low residuals, corresponding to
areas where more points were expected than observed (overprediction),
and brightly shaded blue indicates unusually high residuals,
corresponding to areas of underprediction of seismicity.

The tiles in the Voronoi residual plot in Figure~\ref{correctModel}
range from light to moderate hues, representing residuals that are
within the range expected under the reference distribution. Similarly,
the histogram and quantile plot of the Voronoi residuals demonstrate
that the distribution of the residuals is well approximated by
distribution (\ref{eq2}).

\subsection{Misspecification}\label{sec3.2}
In order to evaluate the ability of Voronoi residuals to detect model
misspecification, simulations were obtained using a generating model
and then residuals were computed based on a different proposed model.
The top left panel of Figure~\ref{overestimating} displays a
realization of a Poisson process with intensity $\lambda(x, y) = 100
\mathbf{1}_{\{\llvert X\rrvert, \llvert Y\rrvert > 0.35\}}$.

The proposed model assumes a constant intensity across the space, $\hat
\lambda= 100$. Because of the lack of points near the origin, the
tiles near the origin are larger than expected under the proposed model
and, hence, for such a cell $C$ near the origin, the integral $\int
_{C} \hat\lambda(x,y) \,\mathrm{d}x \,\mathrm{d}y$ exceeds 1,
leading to negative residuals of large absolute value. These unusually
large negative residuals are evident in the Voronoi residual plot and
clearly highlight the region where the proposed model overpredicts the
intensity of the process. These residuals are also clear outliers in
the left tail of the reference distribution of the residuals and, as a
result, one sees deviations from the identity line in the
quantile--quantile plot in Figure~\ref{overestimating}.

%
%
\section{Statistical power}\label{sec4}\label{secpower}
We now consider the manner in which the statistical power of residual
analysis using a Voronoi partition differs from that of a pixel
partition. In the context of a residual plot, a procedure with low
power would generate what appears to be a structureless residual plot
even when the model is misspecified. To allow for an unambiguous
comparison, here we focus on power in the formal testing setting: the
probability that a misspecified model will be rejected at a given
confidence level.

\subsection{Probability Integral Transform}\label{sec4.1}\label{secPIT}
As was discussed in Section~\ref{secvoronoiresiduals}, the
distribution of Voronoi residuals under the null hypothesis is well
approximated by a modified gamma distribution, while the distribution
of pixel residuals is that of a Poisson distributed variable with
intensity $\int_{G_i} \hat{\lambda} \,\mathrm{d}\mu$, for pixel
$G_i$, centered to have mean zero. To establish a basis to compare the
consistency between proposed models and data for these two methods, we
utilize the Probability Integral Transform (PIT) [\citet{Dawid1984}].
The PIT was proposed to evaluate how well a probabilistic forecast is
calibrated by assessing the distribution of the values that the
observations take under the cumulative distribution function of the
proposed model. If the observations are a random draw from that model,
a histogram of the PIT values should appear to be standard uniform.

One condition for the uniformity of the PIT values is that the proposed
model be continuous. This holds for Voronoi residuals, which are
approximately gamma distributed under the null hypothesis, but not for
the Poisson counts from pixel residuals.
For such discrete random variables, randomized versions of the PIT have
been proposed. Using the formulation in \citet{Czado2009}, if $F$ is the
distribution function of the proposed discrete model, $X \sim F$ is an
observed random count and $V$ is standard uniform and independent of
$X$, then $U$ is standard uniform, where
%
%
\begin{eqnarray}
U &=& F (X-1 ) + V \bigl(F (X ) - F (X-1 ) \bigr),\qquad X \ge1,
\\
U &=& V F (0 ), \qquad X=0. \label{eq4}
\end{eqnarray}
The method can be thought of as transforming a discrete c.d.f.
into a continuous c.d.f. by the addition of uniform random noise.

\subsection{Formal testing}\label{sec4.2}
The PIT, both standard and randomized, provides a formal basis for
testing two competing residual methods. For a given proposed model and
a given realization of points, the histogram of PIT values, $u_1, u_2,\ldots, u_n$, for each residual method should appear standard uniform
if the proposed model is the same as the generating model. The
sensitivity of the histogram to misspecifications in the proposed model
reflects the statistical power of the procedure.

There are many test statistics that could be used to evaluate the
goodness of fit of the standard uniform distribution to the PIT values.
Here we choose to use the Kolmogorov--Smirnov (K--S) statistic [\citet
{Massey1951}],
\[
D_n = \sup_n \bigl\llvert F_n (
x ) - F ( x )\bigr\rrvert,
\]
where $F_n ( x )$ is the empirical c.d.f. of the sample and $F (
x )$ is the c.d.f. of the standard uniform. Since the Voronoi residuals of
a given realization are not independent of one another, we use critical
values from a simulated reference distribution instead of the limiting
distribution of the statistic.

\subsection{Simulation design}\label{sec4.3}
Two models were considered for the simulation study. The first was a
homogeneous Poisson model on the unit square with intensity $\lambda$
on~$\mathbb{R}^2$. The second was an inhomogeneous Poisson model with intensity
%
%
\begin{equation}
\lambda(x,y) = 100 + 200 \bigl( \tilde{x}^\beta\tilde{y}^\beta
c_{\beta} \bigr), \label{modelbeta}
\end{equation}
on $\mathbb{R}^2$, where $\tilde{x} = \frac{1}{2} - \llvert
x - \frac{1}{2}\rrvert $, $\tilde{y} = \frac{1}{2} - \llvert y - \frac
{1}{2}\rrvert $ and $\beta> 0$. The constant $c_{\beta}$ is a
scaling constant chosen so that the parenthetical term integrates to
one. The result is a function that is symmetric about $x=0.5$ and
$y=0.5$, reaches a maximum at $(0.5, 0.5)$, integrates to 300 regardless of
the choice of $\beta$, and is reasonably flat along the boundary box.
This final characteristic should allow the alternative approach to the
boundary problem, described below, to be relatively unbiased.
Additionally, it presents inhomogeneity
similar to what might be expected in an earthquake setting.

The procedure for the inhomogeneous simulation was as follows. A point
pattern was sampled from the true generating model, (\ref{modelbeta})
with $\beta=4$. For a given proposed model, (\ref{modelbeta}) with
$\beta= \beta_0$, and a fixed number of pixels $n$ on the unit square
$[0,1]^2$, PIT values were calculated for the counts in each pixel, $G_i$.
The empirical c.d.f. of the PIT transformed residuals was then compared to
the c.d.f. of the standard uniform using the K--S test. After many
iterations of this procedure, the proportion of iterations with an
observed K--S statistic that exceeded the critical value served as the
estimate of the power of the method. An analogous procedure was
followed for the Voronoi partition, but with the PIT values calculated
by evaluating the Voronoi residuals (\ref{eq1}) under the gamma
distribution (\ref{eq2}).

The homogeneous simulation was conducted in the same manner, but drew
samples from a generating model of $\lambda= 500$ and compared them to
estimates from a proposed model $\lambda_0$.

%
\begin{figure}

\includegraphics{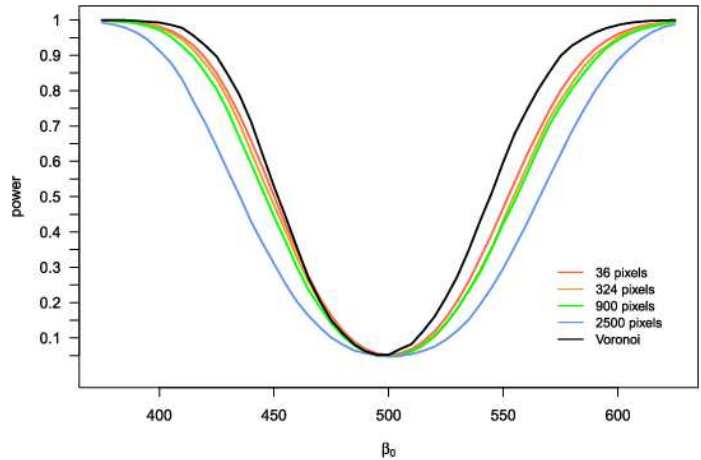}

\caption{Estimated power curves for the K--S test based on five
different pixel partitions as well as the Voronoi tessellation. The
model under consideration is homogeneous Poisson with a generating
intensity of $\lambda= 500$.}
\label{hpow}
\end{figure}

It is known that Voronoi cells generated along the boundary of the
space do not follow the same distribution as the interior cells. One
recourse is to omit them from the analysis, as in Section~\ref{sec3}. Here we
consider realizations of the model (\ref{modelbeta}) on the entire
plane but consider only the distribution of all cells generated by
points inside the unit square $[0,1]^2$.


\subsection{Results}\label{sec4.4}
For the homogeneous model, Figure~\ref{hpow} shows the resulting
estimated power curves for several pixel partitions, including $n \in\{
36, 324,\break  900, 2500\}$. The power of each method was computed for a
series of proposed models, $\lambda_0 \in(375, 625)$.
The best performance was by the method that used the Voronoi partition,
which shows high power throughout the range of misspecification. For
the pixel partitions, $n=36$ had the highest power, but as the number
of partitions increases, the K--S test loses its power to detect
misspecification. This trend can be attributed to characteristic I:
when the space is divided into many small cells, the integrated
conditional intensity is very small and the distribution of the
residuals is highly skewed. As a consequence, the majority of counts
are zeros, so the majority of the PIT values are being generating by $V
F ( 0 )$ [equation~(\ref{eq4})] and, thus, the resulting residuals have
little power to detect model misspecification. The most powerful test
in this homogeneous setting is in fact one with no partitioning, which
is equivalent to the Number-test from the earthquake forecasting
literature [\citet{Schorlemmer2007b}].

\begin{figure}

\includegraphics{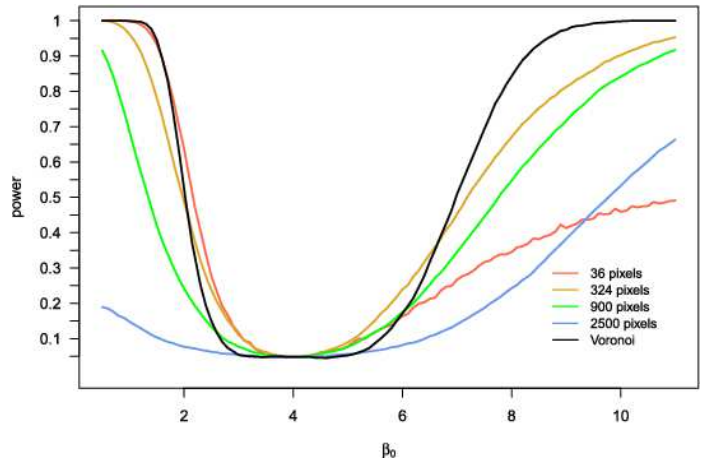}

\caption{Estimated power curves for the K--S test based on five
different pixel partitions as well as the Voronoi tessellation. The
model under consideration is $\lambda(x,y) = 100 + 200 ( \tilde
{x}^\beta\tilde{y}^\beta c )$, where $\tilde{x} = \frac{1}{2} -
\llvert x - \frac{1}{2}\rrvert $ and $\tilde{y} = \frac{1}{2} -
\llvert y - \frac{1}{2}\rrvert $. The generating model is $\beta=4$.}
\label{ih1pow}
\end{figure}

For the inhomogeneous case, power curves were computed for a series of
proposed models of the form (\ref{modelbeta}), with $\beta_0 \in(0.5, 11)$. The
results are shown in Figure~\ref{ih1pow}. The power curve for the
Voronoi method presents good overall performance, particularly when the
model is substantially misspecified. The Voronoi residuals are not
ideally powerful for detecting slight misspecification, however,
perhaps because the partition itself is random, thus introducing some
variation that is difficult to distinguish from a small change in $\beta$.

%

Focusing only on the four pixel methods, the best performance is at
$n=324$ pixels. The poor performance of $n=36$ in detecting the large
positive misspecification is due to the fact that the model becomes
more inhomogeneous as $\beta_0$ increases, but that inhomogeneity is
averaged over cells that are too large (the problem associated with
characteristic II in Section~\ref{sec1}). Meanwhile, the poor overall
performance of $n=2500$ is due to the same problem that exists in the
homogeneous setting, where the PIT values are dominated by the random
uniform noise.

In applications such as earthquake modeling, the use of pixel methods
often results in situations with extremely low intensities in some
pixels, similar to the case considered here with $n=2500$, but perhaps
even more extreme. For instance, one of the most successful forecasts
of California seismicity [\citet{Helmstetter2007}] estimated rates in
each of $n=7682$ pixels in a model that estimated a total of only 35.4
earthquakes above M 4.95 over the course of a prediction experiment
that lasted from 2006 to 2011. Estimated integrated rates were as low
as 0.000007 in some pixels, and 58\% of the pixels had integrated rates
that were lower than 0.001. An immediate improvement could be made by
aggregating the pixels, but this in turn will average over the strong
inhomogeneity along fault lines in the model, which will lower power.
For this reason, the Voronoi residual method may be better suited to
the evaluation of seismicity models, as well as other processes that
are thought to be highly inhomogeneous.

%
%
\section{Application to models for Southern California seismicity}\label{sec5}

\subsection{The ETAS model and the Hector Mine earthquake catalog}\label{sec5.1}
In this section we apply Voronoi residual analysis to the
spatial--temporal Epidemic-Type Aftershock Sequence (ETAS) model of
\citet{Ogata1998}, which has been widely used to describe
earthquake catalogs.

According to the ETAS model of \citet{Ogata1998}, the conditional
intensity $\lambda$ may be written
%
\begin{equation}
\lambda(t,x,y) = \mu\rho(x,y) + \sum_{j\dvtx t_j < t}
g(t-t_j, x-x_j, y-y_j; M_j),
\label{eq3}
\end{equation}
where $\rho(x,y)$ is a spatial density on the spatial observation
region $S$,
$t$ and $(x,y)$ are temporal and spatial coordinates, respectively,
$M_j$ is the magnitude of earthquake $j$, and
where the triggering function, $g$, may be given by one of several
different forms. One form for $g$ proposed in \citet{Ogata1998} is
%
\begin{equation}
g(t,x,y; M) = K (t+c)^{-p} e^{a(M-M_0)} \bigl(x^2 +
y^2 + d\bigr)^{-q},
\end{equation}
%
where $M_0$ is the lower magnitude cutoff for the catalog.

\begin{figure}

\includegraphics{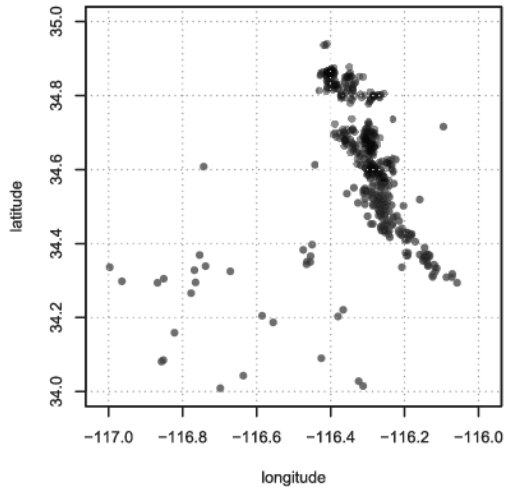}

\caption{Locations of the 520 earthquakes that make up the Hector Mine
earthquake sequence. All events occurred between 10/16/1999 to
12/23/2000 in the Mojave Desert 35 miles east of Barstow, CA and were
above magnitude 3.0.}
\label{hector-pattern}
\end{figure}

There is considerable debate in the seismological community about the
best method to estimate the spatial background rate $\rho(x,y)$ [\citet{Ogata2011,Helmstetter2012,Zhuang2012}]. When modeling larger, regional
catalogs, $\rho$ is often estimated by smoothing the largest events in
the historical catalog [\citet{Ogata1998}], and in such cases a very
important open question is how (and how much) to smooth [\citet
{Schoenberg2003,Helmstetter2007,Helmstetter2012,Zhuang2012}]. For a
single earthquake-aftershock sequence, however, can one instead simply
estimate $\rho$ as constant within a finite, local area, as in
\citet{Schoenberg2013}? A prime catalog to investigate these
questions is the catalog of California earthquakes including and just
after the 1999 Hector Mine earthquake (Figure~\ref{hector-pattern}).
This data set was analyzed previously in \citet{Ogata2003}, and
consists of the origin times, epicentral locations and magnitudes of
the 520 earthquakes with magnitude at least $3.0$, from latitude 34 to
35, longitude $-$116 to $-$117, from 10/16/1999 to 12/23/2000, obtained
from the Southern California Seismic Network (SCSN).

%

The parameters in the model were estimated by maximum likelihood
estimation, using the progressive approximation technique described in
\citet{Schoenberg2013}. For the purpose of this analysis, we
focused on the purely spatial aspects of the residuals, and thus
integrated over the temporal domain to enable planar visualization of
the residuals. The result is a Voronoi tessellation of the spatial
domain where for tile $C_i$, for the integral in equation (\ref{eq1}), the
estimated conditional intensity function $\hat\lambda(t,x,y)$ is
numerically integrated over the spatial tile $C_i$ and over the entire
time domain from 10/16/1999 to 12/23/2000.

\subsection{Assessing the fit of the model}\label{sec5.2}
We take two approaches to detecting inconsistencies between the ETAS
model and the Hector Mine catalog: the inspection of plots for signs of
spatial structure in the residuals and the evaluation of PIT histograms
as overall indicators of goodness of fit.

%
\begin{figure}

\includegraphics{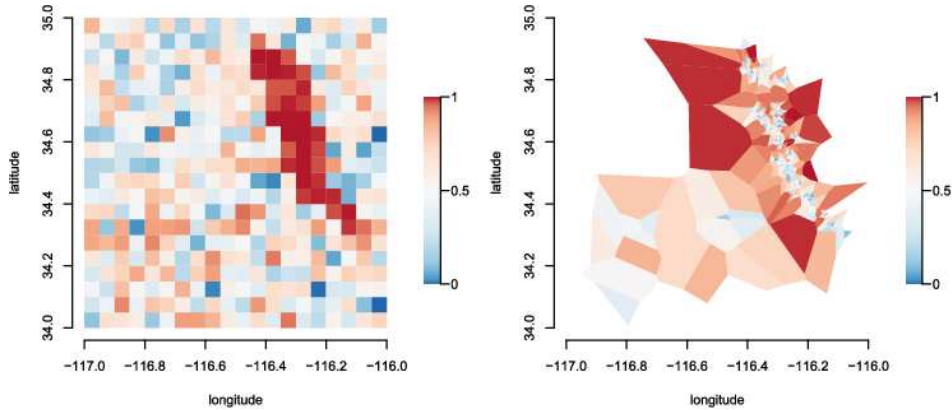}

\caption{Residual plots of fitting the ETAS model to the Hector Mine
sequence using the pixel (left panel) and Voronoi partition (right
panel). In both plots, the PIT value is transformed to a color scale
using the inverse normal transformation. In the Voronoi plot, tiles
intersecting the boundary of the space are ignored, as the distribution
of these tile areas may differ substantially from the gamma distribution.}
\label{hector-res-plots}
\end{figure}

Figure~\ref{hector-res-plots} shows residual plots based on both the
pixel partition and the Voronoi tessellation. As in Figures~\ref
{correctModel} and~\ref{overestimating}, the magnitude of each residual
cell is represented by the value that the residual takes under the
distribution function appropriate to that model (Poisson or gamma),
which is the PIT value discussed in Section~\ref{secPIT}. The pixel
residual plot shows that the ETAS model estimates a much higher
conditional intensity along the fault region running from ($-$116.4,
33.9) to ($-$116.1, 33.3) than was observed in the Hector Mine sequence.
Away from the fault, the residuals are less structured, with no
indication of model misspecification.

\begin{figure}

\includegraphics{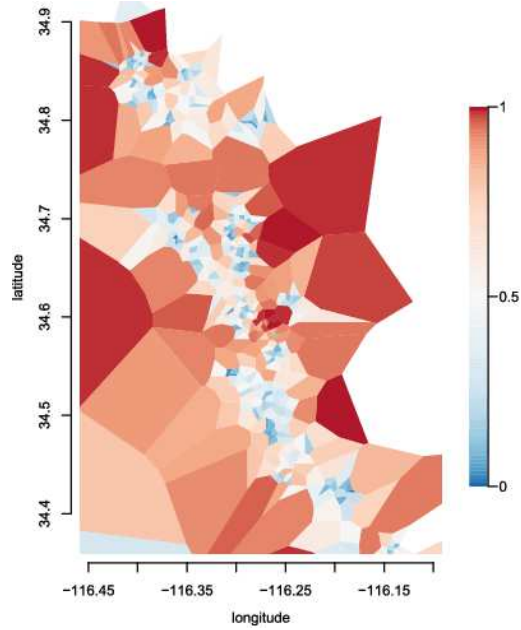}

\caption{Enlarged Voronoi residual plot focusing on the region of
Figure~\protect\ref{hector-res-plots} that covers the fault line, which
runs from approximately ($-$116.4, 34.85) to ($-$116.25, 34.4). PIT values
are transformed to a color scale using the inverse normal
transformation and tiles intersecting the boundary of the space are
ignored, as the distribution of these tile areas may differ
substantially from the gamma distribution.}
\label{hector-res-plot-zoom}
\end{figure}

The Voronoi residual plot shares the same color scale as the pixel
plot, but excludes the boundary cells by shading them white. Some
strong overprediction is apparent in several large cells in the general
area of the fault line, but the structure is more nuanced than that
found in the pixel plot. Figure~\ref{hector-res-plot-zoom} provides an
enlarged version of the fault region, showing systematic
underprediction along the fault and overprediction on the periphery of
the fault. Such structure in the residuals indicates that the ETAS
model with uniform background rate may be oversmoothing. This suggests
modeling the background rate $\rho(x,y)$ in equation (\ref{eq3}) as
inhomogeneous for Southern California seismicity, in agreement with
\citet{Ogata1998} who came to a similar conclusion for Japanese seismicity.

%

This structure is lost when the residuals are visualized using the
pixel partition because the over- and underprediction are averaged over
the larger fixed cells (a~case of characteristic II). The true
intensity in this region is likely highly spatially variable, which
makes the spatially adaptive Voronoi partition a more appropriate choice.

%
\begin{figure}

\includegraphics{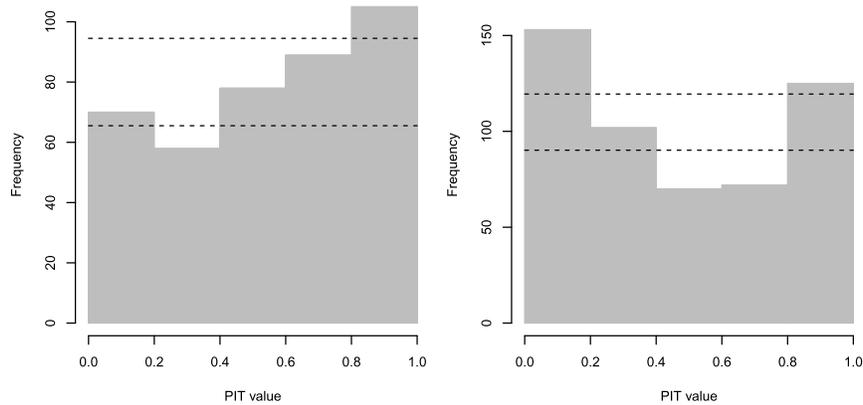}

\caption{Histograms of PIT values from the ETAS model and Hector Mine
catalog based on the pixel partition (left panel) and the Voronoi
tessellation (right panel). Dashed lines represent pointwise 90\%
coverage intervals calculated by simulation.}
\label{hector-pit-hists}
\end{figure}

As discussed in Section~\ref{secPIT}, PIT values will be uniformly
distributed if the fitted model is correct, therefore, PIT histograms
can be used as a means to assess general goodness of fit [\citet
{Thorarinsdottir2013}]. Figure~\ref{hector-pit-hists} shows the
distribution of the randomized PIT values resulting from the pixel
partition (left panel) alongside the PIT values from the Voronoi
partition (right panel). Both histograms show deviations from
uniformity, suggesting model misspecification. The histogram resulting
from the Voronoi parition suggests more deviation, however, which is
consistent with the finding in Section~\ref{secpower} that this
partition is more sensitive to misspecification than the pixel
partition. It also suggests that there are areas of strong
underprediction as well as overprediction, while the pixel PIT values
primarily identify the overprediction. The PIT histogram is a useful
tool to visualize overall goodness of fit, while the Voronoi residual
plot seems to be more powerful for identifying areas of poor fit.

\section{Discussion}\label{sec6}
Applying Voronoi residual analysis to the ETAS model and the Hector
Mine earthquake sequence suggests model misspecification---over\-smoothing along the fault---that is undetected by other methods.
These Voronoi residuals may of course be used in tandem with standard,
pixel-based residuals, which may in turn be based on a judicious choice
of pixel size, or perhaps using a different spatially adaptive grid
than the one proposed here.

The use of PIT values, both in residual plots and histograms, relies
upon a readily computable form for $F$, the distribution of residuals
under the fitted model. In the case of the Voronoi partition, this
requires Monte Carlo integration of the conditional intensity function
over the irregular cells. This process can be time consuming if the
intensity function is sufficiently inhomogenous or if the number of
earthquakes in the catalog is very high. The PIT values of the pixel
partition are easier to compute and they benefit from a more
straightforward interpretation in the residual plot simply because the
fixed grid is a more familiar configuration. However, because of their
improved statistical power, Voronoi residuals are more informative and
thus worth the additional computation and consideration.

The importance of selecting the size of the cell on which to compute a
residual is not unique to this PIT--K--S statistic testing environment.
The discrepancy measure proposed by \citet{Guan2008} is defined on
a Borel set of a given shape $S$. The author emphasizes the importance
of choosing an appropriate size for $S$ (page~835) and points out that
if the cell is too small or too large, the power will suffer. A related
problem arises in the selection of the bandwidth of the kernel used to
smooth a residual field [\citet{Baddeley2005b}, Section~13 and discussion].

Although we have focused on formal testing at the level of the entire
collection of residuals, testing could also be performed at the level
of individual cells. For the Voronoi partition, this extension is
straightforward and is essentially what is being done informally in the
shaded residual plots. For any pixel partition, such testing may be
problematic, as any pixel with an integrated conditional intensity
close to zero would contain zero points with more than $95\%$
probability, so any hypothesis test with $\alpha= 0.05$ using a
rejection interval would necessarily have a type I error near 1.

Generating the partition using a tessellation of the observed pattern
has advantages and disadvantages. The advantage is that it is adaptive
and requires no input from the user regarding tuning parameters. The
disadvantages are that some sampling variability is induced by the
random cell areas and that the residuals are dependent, so techniques
relying upon an i.i.d. assumption must be used cautiously. A promising
future direction is to consider residuals based on a model-based
centroidal Voronoi tessellation [\citet{Du1999}], which mitigates
characteristics I and II of the pixel method while providing a
partition that creates residuals that are independent of one another if
the underlying model is Poisson.

It should also be noted that the standardization methods proposed in
\citet{Baddeley2005b} may be used with Voronoi residuals or
instead one may elect to plot deviance residuals [\citet{Clements2011}]
in each of the Voronoi cells. In general, our experience suggests that
the standardization chosen for the residuals seems far less critical
than the choice of grid. The results seem roughly analogous to kernel
density estimation, where the selection of a kernel function is far
less critical than the choice of bandwidth governing its range.

\section*{Acknowledgments}
We thank the reviewer and Associate Editor for helpful comments that
significantly improved this paper.


%

\printaddresses
\end{document}